\documentclass{article}
\usepackage{spconf}
\usepackage{cite}
\usepackage{amsmath,amssymb,amsfonts}
\usepackage{algorithmic}
\usepackage{graphicx}
\usepackage{textcomp}
\usepackage{booktabs}
\usepackage{xcolor}
\usepackage{hyperref}
\usepackage{multicol}
\usepackage{multirow}

\usepackage{multirow}

\title{SongPrep: A Preprocessing Framework and end-to-end model for full-song structure parsing and lyrics transcription}
\name{
    \begin{tabular}[c]{@{}c@{}}
        Wei Tan\thanks{
            Demos can be found inhttps://song-prep.github.io/demo/ \\
            Data can be found in https://huggingface.co/datasets/waytan22/SSLD-200
        }, Shun Lei,
        Huaicheng Zhang, Guangzheng Li, 
        Yixuan Zhang, \\
        Hangting Chen, 
        Jianwei Yu, Rongzhi Gu, Dong Yu
    \end{tabular}
}

\address{Tencent AI Lab}

\begin{document}
\maketitle
\begin{abstract}
Artificial Intelligence Generated Content (AIGC) is currently a popular research area. Among its various branches, song generation has attracted growing interest. Despite the abundance of available songs, effective data preparation remains a significant challenge. Converting these songs into training-ready datasets typically requires extensive manual labeling, which is both time consuming and costly. To address this issue, we propose SongPrep, an automated preprocessing pipeline designed specifically for song data. This framework streamlines key processes such as source separation, structure analysis, and lyric recognition, producing structured data that can be directly used to train song generation models. Furthermore, we introduce SongPrepE2E, an end-to-end structured lyrics recognition model based on pretrained language models. Without the need for additional source separation, SongPrepE2E is able to analyze the structure and lyrics of entire songs and provide precise timestamps. By leveraging context from the whole song alongside pretrained semantic knowledge, SongPrepE2E achieves low Diarization Error Rate (DER) and Word Error Rate (WER) on the proposed SSLD-200 dataset. Downstream tasks demonstrate that training song generation models with the data output by SongPrepE2E enables the generated songs to closely resemble those produced by humans.
\end{abstract}
\begin{keywords}
Song generation, data processing, structure analysis, lyric recognition
\end{keywords}

\section{Introduction}
In recent years, with the rapid development of deep learning, AIGC models have shown great potential to generate high-quality audio and images. However, collecting high-quality large-scale training data with accurate annotations remains a critical factor in achieving high-quality generation. Over the past few years, the research community has invested substantial time and effort in building datasets that have significantly advanced technologies such as automatic speech recognition (ASR), text-to-speech synthesis (TTS), and image generation. 

In contrast, music generation, particularly song generation with vocals, still suffers from limited data resources. Existing datasets\cite{BertinMahieux2011TheMS, Zhang2024GTSingerAG} often lack structured lyrics or are restricted in scale, making them difficult to use directly for training song generation models\cite{lei2025levo,yang2025songbloom}. It has become a major bottleneck for further progress in song generation.

Leveraging the vast amount of freely available song data offers a potential solution\cite{Yu2023AutoPrepAA}. But most open song data cannot be used directly because it lacks accurate annotations, such as lyric and structural information. Training models with incomplete or incorrect lyrics can cause severe hallucinations\cite{zhang2025hallucinationfreemusicreinforcementlearning}, while missing structure makes it difficult for models to learn melodies that match musical theory. To address these challenges, we propose an automatic processing framework for song data, named SongPrep. The framework integrates several models and tools, including Demucs\cite{Rouard2022HybridTF} for source separation, All-In-One\cite{Kim2023AllinOneMA} for structure parsing, and a Zipformer-based ASR model\cite{Yao2023ZipformerAF} for lyric transcription. Furthermore, we introduce an end-to-end model for full song structure parsing and lyrics transcription, named SongPrepE2E, which combines MuCodec \cite{zhu2025muq} with a pretrained large language model (LLM). SongPrepE2E is trained on the data produced by SongPrep, allowing it to generate structured lyrics directly from songs in an end-to-end manner.

Due to the lack of annotations, it is difficult to directly evaluate annotation performance. Therefore, we construct a test dataset called SSLD-200 (Song Structure and Lyric Dataset), which includes 100 Chinese songs and 100 English songs, along with detailed structural information and lyrics. Using this dataset, we conduct comprehensive experiments to assess the performance of the models mentioned above.

The main contributions of this paper are summarized as follows:
1) We propose SongPrep, one of the first frameworks capable of automatically generating structural information and lyrics for a diverse range of songs.
2) We develop SongPrepE2E, an end-to-end structured lyrics recognition model that achieves better performance than the multistage SongPrep pipeline.
3) We release SSLD-200, an open source dataset comprising 200 songs with detailed manual annotations, establishing a benchmark for the evaluation of structured lyrics recognition in the future.

\begin{figure*}[htbp]
\includegraphics[width=2.0\columnwidth]{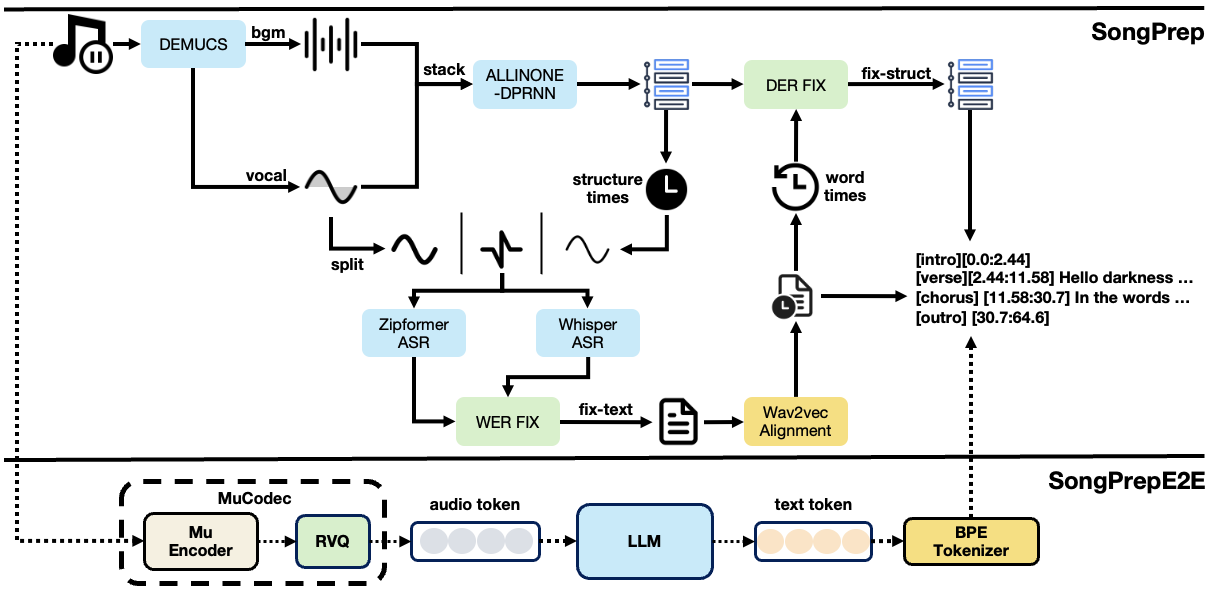}
\caption{The diagram of the proposed SongPrep and SongPrepE2E framework. The models output in the format [structure][start:end]lyric, where structure is the label from StructureAnalysis for the segment, start and end are the segment’s start and end times, and lyric is the recognized lyrics. The SongPrep additionally computes WER based on a dual-head ASR.}
\label{fig1}
\end{figure*}

\section{Method}
The upper part of Figure 1 illustrates the workflow of the proposed SongPrep framework. Raw song data are sequentially processed through source separation, structure analysis, and lyric recognition, ultimately transforming them into structured data.

Since a song typically consists of multiple tracks, each analysis module uses only the subset of tracks relevant to its specific task. Specifically, we employ the Demucs model for source separation, decomposing the song into four tracks: vocals, drums, bass, and other instruments. The four separated tracks are reassembled for structure prediction, while the vocal track is processed by an ASR model to extract the lyrics.

\subsection{Structure Analysis}
To enable the model to learn musicality during training and control song generation using structural information during inference, our pipeline must accurately analyze the structural attributes of songs. For this purpose, we employ the All-In-One model, which enhances CNN architecture with one-dimensional and two-dimensional dilated neighborhood attention (1D NA and 2D NA). However, the structural information produced by the All-In-One model in our pipeline did not fully meet expectations, motivating the following improvements.

Firstly, the All-In-One model was initially trained solely on English songs, resulting in suboptimal performance when applied to Chinese songs. To address this limitation, we constructed a bilingual training dataset comprising 3,700 songs sourced from internal copyright materials for retraining. Secondly, the original label set included numerous categories, some of which were ambiguous or unsuitable for song generation purposes. We therefore refined the label set from 10 to 7 categories: intro, outro, inst, verse, chorus, bridge, and silence. Thirdly, while dilated neighborhood attention is effective in capturing local semantic information, it falls short in modeling global song structure. To compensate, we inserted a Dual-Path RNN (DPRNN) \cite{Luo2019DualPathRE} block after every three All-In-One blocks, as shown in Figure 2, providing the model with improved capacity for learning global semantic dependencies.

Together, these three modifications reduced the model's DER from 25.0\% to 16.1\% on our proposed dataset.

\begin{figure}[htbp]
\includegraphics[width=1.0\columnwidth]{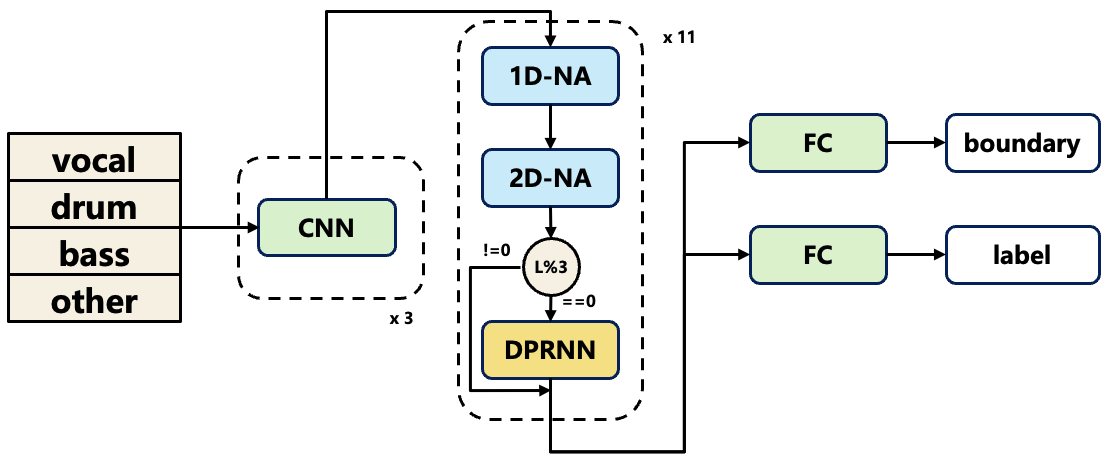}
\caption{Structure of All-In-One DPRNN}
\label{fig2}
\end{figure}

\subsection{Lyric Recognition}
After structural analysis, we apply an ASR system to perform lyric transcription on the vocal track. Specifically, we process only the vocal segments within the verse, chorus, and bridge sections, as these are the primary structures containing vocal performances in music theory. The lyric recognition workflow consists of the following stages.

Firstly, we obtain the audio of the song and the corresponding lyrics through web scraping. However, the quality of these lyric texts does not meet the training requirements for missing, mismatching, and misalignment. To address these issues, we employ Whisper \cite{radford2022robustspeechrecognitionlargescale} for lyric recognition and apply an improved WER-FIX algorithm to verify and optimize lyrics. WER-FIX consists of two steps. First, we retain only those lyrics with a WER below 0.7. Second, we further refine the retained lyrics using WER scoring. In the second process, we assume that both the characters in the original lyrics and the number of characters predicted by the ASR system are generally accurate. For substitution errors identified by WER, we adopt the results from the original lyrics, whereas for insertion and deletion errors, we rely on the ASR outputs.

Secondly, we utilize the validated lyric data from the previous step to construct a fine-tuning dataset comprising approximately 100,000 songs and totaling around 7,000 hours. This dataset is used to fine-tune a Zipformer-based ASR system. In the proposed dataset, the Zipformer ASR achieves a WER of 25.8\%, outperforming Whisper, which obtains a WER of 27.7\%. By integrating both ASR systems, we establish a dual-head ASR framework capable of processing song audio in cases where annotated lyrics are unavailable.

Finally, we incorporate a word alignment module based on wav2vec2\cite{Baevski2020wav2vec2A} to calibrate the results of the structural analysis. This alignment module helps prevent large sections of instrumental accompaniment from being incorrectly labeled as verse or chorus due to errors in the structural analysis.


\subsection{End to End model For Structured Lyrics Recognition}
While the proposed SongPrep framework achieves improvements in both structure parsing and lyric transcription, it still suffers from two major limitations. First, the multistage pipeline leads to low inference efficiency. Second, ASR frameworks usually require splitting long audio into short chunks, which causes the loss of contextual and semantic information, thereby reducing the recognition accuracy. To address these challenges, we integrate MuCodec with LLM to propose SongPrepE2E, an end-to-end system trained on the data curated by SongPrep. SongPrepE2E is capable of directly extracting structured lyric transcriptions from full-length songs, achieving better recognition accuracy and deployment efficiency.

Initially, we discretize the audio into tokens using MuCodec, which consists of MuEncoder\cite{zhu2025muq}, Residual Vector Quantization (RVQ) and flow matching module\cite{lipman2023flowmatchinggenerativemodeling}. The MuEncoder utilizes 13 stacked Conformer\cite{gulati2020conformer} blocks and is trained with multiple constraints, including Masked Language Modeling (MLM), reconstruction, and lyric recognition. After training, we freeze the MuEncoder and jointly train the RVQ and flow matching modules, optimizing for reconstruction loss, codebook loss, and representation alignment. Representation alignment ensures that intermediate representations of the flow matching model match the Hubert\cite{hsu2021hubertselfsupervisedspeechrepresentation} features of the audio, guiding the token to capture more semantic information. The resulting MuCodec produces an audio token sequence at 25 Hz with a codebook size of 16,384.

Next, we pair the audio tokens generated by MuCodec with SongPrep's structured lyrics output to create audio–text training pairs. These pairs are then used to perform Supervised Fine-Tuning (SFT) of Qwen2-7B\cite{yang2024qwen2technicalreport}. After training, SongPrepE2E can process a complete song up to four minutes in duration and directly output its structured lyrics, indicated by the lower part of Figure 1. As demonstrated by the experiments in Section 3, SongPrepE2E achieves better performance compared to SongPrep. Moreover, this training approach may generalize well to other large language model architectures and weights.

\section{Experiments}
To evaluate the performance of the proposed systems, we apply them to our proposed SSLD-200 dataset. Furthermore, to assess their practical effectiveness, we utilize the processed output in downstream song generation tasks.

\subsection{Dataset}
To provide a more intuitive assessment of the architecture’s performance, we introduce a validation set named SSLD-200. SSLD-200 consists of 200 songs, 100 English and 100 Chinese, collected entirely from YouTube, with a total duration of 13.9 hours. For each song, we segment the audio based on structure labels as described in Section 2.1, and annotate the boundaries at the level of seconds. Additionally, for each segment, we provide the corresponding lyric text for the audio within its annotated time range.

\subsection{Performance of each module}
Table 1 presents the processing results for each module in SongPrep on the SSLD-200. The following key observations were made across different modules.

We evaluate the structure analysis module using the DER. Because DER can simultaneously represent the accuracy of both the labels and their corresponding time intervals. The ALL-IN-ONE model, fine-tuned with 3,700 annotated songs, achieves a DER reduction of 4.2\% compared to the original model. Further improvements are observed after integrating the DPRNN module, which incorporates a global contextual perspective, leading to a further reduction in DER to 16.1\%. These results suggest that the main performance bottleneck remains the quantity of training data. We anticipate that increasing the size of the training dataset will yield even greater performance gains.

The lyric recognition module is assessed with the WER metric. Using clean vocal separation tracks for ASR significantly enhances Whisper’s performance, improving WER from 47.2\% to 27.7\%. Furthermore, the Zipformer base ASR model, fine-tuned on data corrected by the WER FIX algorithm, achieves an even lower WER of 25.8\%.

\begin{table}[!h]
\label{tab:table1}
\centering
\caption{Performance of each module}
\begin{tabular}{c|c|c|c}
\toprule[2pt]
\textbf{Module} & \textbf{Metrics} & \textbf{Method} & \textbf{Value} \\ 
\hline
\multirow{3}{*}{\begin{tabular}[c]{@{}c@{}}Structure\\ Analysis\end{tabular}}
 & \multirow{3}{*}{\begin{tabular}[c]{@{}c@{}}DER $\downarrow$ \end{tabular}} 
 & ALL-IN-ONE & 25.0\% \\
& & +Fine Tune & 20.8\% \\
& & +DPRNN & \textbf{16.1\%} \\
\hline
\multirow{4}{*}{\begin{tabular}[c]{@{}c@{}}Lyric\\ Recognition\end{tabular}}
 & \multirow{4}{*}{\begin{tabular}[c]{@{}c@{}}WER $\downarrow$ \end{tabular}}
 & Whisper ASR & 47.2\% \\
&  & +Demucs & 27.7\% \\
&  & Zipformer ASR & 30.6\% \\
&  & +Demucs & \textbf{25.8\%} \\
\bottomrule[2pt]
\end{tabular}
\centering
\end{table}


\subsection{Processing results of pipeline}
This section presents the overall performance of SongPrep and SongPrepE2E. We first evaluate SongPrep by connecting its modules sequentially, with the results summarized in Table 2. The findings indicate that the performances of the individual modules are interdependent. For example, correcting the Structure Analysis results using alignment from the ASR module reduces the DER from 16.1\% to 15.8\%. However, errors originating in the Structure Analysis step can propagate and adversely affect the Lyric Recognition module, increasing the WER from 25.8\% to 27.7\%.

Using an in-house dataset of 2 million songs (approximately 110,000 hours), cleaned by SongPrep, we trained SongPrepE2E by initializing the LLM with pretrained weights. Table 2 also reports the performance of SongPrepE2E under different data filtering strategies. Models trained with a WER \textless 0.3 filter (corresponding to 640 billion tokens) consistently outperformed those filtered with WER \textless 0.1 (240 billion tokens). This is likely due to the stricter WER \textless 0.1 filter yielding a much smaller dataset, which increases the risk of overfitting. Moreover, SongPrepE2E achieves a lower Real Time Factor (RTF) and higher text recognition accuracy compared to SongPrep, which can be attributed to the pretrained language capabilities of the large language model.

\begin{table}[!h]
\label{tab:table2}
\centering
\caption{Performance of the pipeline}
\begin{tabular}{c|c|c|c|c}
\toprule[2pt]
\textbf{Model}  & \textbf{Data Filter} & \textbf{DER} $\downarrow$ & \textbf{WER} $\downarrow$ & \textbf{RTF} $\downarrow$\\ 
\hline
SongPrep  & - & \textbf{15.8\%} & 27.7\% & 0.235 \\ 
\hline
SongPrepE2E  & wer \textless 0.3 & 18.1\% & \textbf{24.3\%} & \textbf{0.108} \\
\hline
SongPrepE2E  & wer \textless 0.1 & 18.8\% & 27.3\% & \textbf{0.108} \\
\bottomrule[2pt]
\end{tabular}
\centering
\end{table}

\subsection{Evaluation on song generation}
To further validate the effectiveness of SongPrepE2E, we train the Levo model \cite{lei2025levo} using internal copyright data that had been processed by SongPrepE2E. Since the original data cannot be used directly for training, we establish a control group using data processed by the All-In-One model along with Whisper, named Base Pipeline, and an experimental group using data processed by SongPrepE2E. Both models are trained using identical architectures.

After training, we generate 40 songs with each model and conduct a subjective evaluation using questionnaires. The questionnaire included three scoring criteria: Musicality Structure, Lyric Matching Degree, and Subjective Bias, each rated on a scale from 1 (“not good at all”) to 5 (“very good”). Thirty participants with basic musical knowledge participate in the test. Detailed results are provided in Table 3.

From Table 3, the following can be observed: 1) In the category of Musicality Structure, SongPrepE2E scores 0.48 points higher than the Base Pipeline. This improvement can be attributed to the Structure Analysis module, suggesting that more accurate structural information contributes to improving control during song generation. 2) In the Lyric Matching Degree category, SongPrepE2E outperforms the Base Pipeline by 1.70 points, owing to the Lyric Recognition module. These results highlight that more accurate text-audio pairs can significantly reduce model hallucination. 3) The score for SongPrepE2E is 0.66 points higher than that for the Base Pipeline, indicating that even with identical data sources, the data processed by SongPrepE2E enables the model to generate more appealing music.

\begin{table}[h]
\label{tab:table5}
\vspace{-4mm}
\caption{Subjective test results of generative model}
\centering
\begin{tabular}{c|c|c}
\toprule[2pt]
\textbf{Metrics} & \textbf{Base Pipeline} & \textbf{SongPrepE2E}  \\ \hline
Musicality Structure$\uparrow$ & 2.52 & \textbf{3.00} \\
Lyric Matching Degree$\uparrow$ & 2.82 & \textbf{4.52} \\
Subjective Bias$\uparrow$ & 2.07 & \textbf{2.73} \\
\bottomrule[2pt]
\end{tabular}
\end{table}

\section{Conclusion}
This paper introduces SongPrep, a song data preprocessing framework capable of directly extracting structural information and lyric from raw song audio. Using data processed by SongPrep, we further develop SongPrepE2E, an end-to-end model for full song structure parsing and lyrics transcription. Experiments on the SSLD-200 benchmark demonstrate that SongPrep and SongPrepE2E can generate high-quality data that can be used directly to train song generation models.

\vfill\pagebreak

\bibliographystyle{IEEEbib}
\bibliography{strings,refs}

\end{document}